\documentclass[useAMS,usenatbib]{mn2e}
\usepackage{graphics,graphicx,array,amssymb,times,fontenc,mathptm}

\title[Variable Stars in the Magellanic Clouds]{Variable Stars in the Magellanic Clouds: \\ Results from OGLE and SIRIUS}
\author[Y. Ita et al.]{Yoshifusa Ita$^{1}$\thanks{E-mail: yita@ioa.s.u-tokyo.ac.jp}, Toshihiko Tanab\'{e}$^{1}$, Noriyuki Matsunaga$^{1}$, Yasushi Nakajima$^{2}$, Chie
\newauthor Nagashima$^{2}$, Takahiro Nagayama$^{2}$, Daisuke Kato$^{2}$, Mikio Kurita$^{2}$, Tetsuya Nagata$^{2}$,
\newauthor Shuji Sato$^{2}$, Motohide Tamura$^{3}$, Hidehiko Nakaya$^{4}$ and Yoshikazu Nakada$^{1,5}$
\\
$^1$Institute of Astronomy, School of Science, The University of Tokyo, Mitaka, Tokyo 181-0015, Japan\\
$^2$Department of Astrophysics, Nagoya University, Chikusa-ku, Nagoya 464-8602, Japan\\
$^3$National Astronomical Observatory of Japan, Mitaka, Tokyo 181-8588, Japan\\
$^4$Subaru Telescope, National Astronomical Observatory of Japan, 650 North A'ohoku Place, Hilo, HI 96720, U.S.A.\\
$^5$Kiso Observatory, School of Science, The University of Tokyo, Mitake, Kiso, Nagano 397-0101, Japan
}
\begin{document}

\date{Received -- / Accepted --}

\pagerange{\pageref{firstpage}--\pageref{lastpage}} \pubyear{2002}

\maketitle

\label{firstpage}

\begin{abstract}We have performed a cross-identification between OGLE-II data and single-epoch SIRIUS near-infrared (NIR) $JHK$ survey data in the Large and Small Magellanic Clouds (LMC and SMC, respectively). After eliminating obvious spurious variables, variables with too few good data and variables that seems to have periods longer than the available baseline of the OGLE-II data, we determined the pulsation periods for 9,681 and 2,927 variables in the LMC and SMC, respectively.

Based on these homogeneous data, we studied the pulsation properties and metallicity effects on period-$K$ magnitude ($PK$) relations by comparing the variable stars in the Large and Small Magellanic Clouds. The sample analyzed here is much larger than the previous studies, and we found the following new features in the $PK$ diagram: (1) variable red giants in the SMC form parallel sequences on the $PK$ plane, just like those found by \citet{wood2000} in the LMC; (2) both of the sequences $A$ and $B$ of \citet{wood2000} have discontinuities, and they occur at the $K$-band luminosity of the TRGB; (3) the sequence $B$ of \citet{wood2000} separates into three independent sequences $B^\pm$ and $C'$; (4) comparison between the theoretical pulsation models (\citealt{wood1996}) and observational data suggests that the variable red giants on sequences $C$ and newly discovered $C^\prime$ are pulsating in the fundamental and first overtone mode, respectively; (5) the theory can not explain the pulsation mode of sequences $A^\pm$ and $B^\pm$, and they are unlikely to be the sequences for the first and second overtone pulsators, as was previously suggested; (6) the zero points of $PK$ relations of Cepheids in the metal deficient SMC are fainter than those of LMC ones by $\approx 0.1$ mag but those of SMC Miras are brighter than those of LMC ones by $\approx 0.13$ mag (adopting the distance modulus offset between the LMC and SMC to be 0.49 mag and assuming the slopes of the $PK$ relations are the same in the two galaxies), which are probably due to metallicity effects.
\end{abstract}


\section{Introduction}
The Large and Small Magellanic Clouds offer us the opportunity to study the stellar evolution. They are close enough to resolve each stellar populations with relatively small telescopes and yet far enough to neglect their depth, so that we can reasonably consider that all stars in them are at the same distance from us.

Many observations have been carried out toward the LMC and SMC, especially for the variable stars to study the relationship between the pulsation period and the luminosity ($PL$ relationship). There is a long standing question related to the $PL$ relationship, whether it depends on metallicity or not. A controversy still exists between observations and theories (e.g., \citealt{alibert} and references therein for Cepheid variables; \citealt{wood1996}, \citealt{glass1995}, \citealt*{feast2002} for long-period variables). Because the $PL$ relationship is of fundamental importance for the determination of the extragalactic distance scale, a homogeneous survey of variable stars in different environments is needed.

Thanks to the gravitational microlensing search projects (OGLE, e.g., \citealt*{udalski}; MACHO, e.g., \citealt{alcock}; EROS, e.g., \citealt{afonso}; MOA, e.g., \citealt{bond}), a large number of variable stars have been found in the Magellanic Clouds. \citet{wood1999} analyzed the data from MACHO survey in the 0.25 square degree area of the LMC bar and showed that the variable red giants form parallel sequences in the period-magnitude plane. \citet{cioni2001} and \citet*{lebzelter} confirmed the results of \citet{wood2000} using data from EROS microlensing survey and AGAPEROS variable star catalog (\citealt*{melchior}), respectively, both in the 0.25 square degree area of the LMC bar.

In this paper, we study the OGLE variables in the 3 square degrees along the LMC bar (see figure 1 of \citealt{ita}) and in the 1 square degree area of the central part of the SMC (figure~\ref{smc}). The OGLE variables have been cross-identified with the SIRIUS NIR survey data and their pulsation periods were determined by applying the Phase Dispersion Minimization technique to the OGLE data. We classify the variable stars on the $PK$ plane, and discuss the properties of each group. The environmental effect on $PK$ relations is discussed by comparing variable stars in the Large and Small Magellanic Clouds.

\section{DATA}
\subsection{IRSF/SIRIUS}
We are now conducting a NIR $JHK$ monitoring survey toward the Magellanic Clouds using the InfraRed Survey Facility (IRSF) at Sutherland, South African Astronomical Observatory. The IRSF consists of a specially constructed 1.4m alt-azimuth telescope to which is attached a large-format 3-channel infrared camera, SIRIUS (Simultaneous three-colour InfraRed Imager for Unbiased Surveys).The SIRIUS can observe the sky in $J$ $H$ and $K_s$ bands simultaneously and has a field of view of about 7.7 arcmin square with a scale of 0.45 arcsec/pixel. Details of the instrument are found in \citet{nagashima} and \citet{nagayama}.

\begin{figure}
\centering
\includegraphics[angle=-90,scale=0.543]{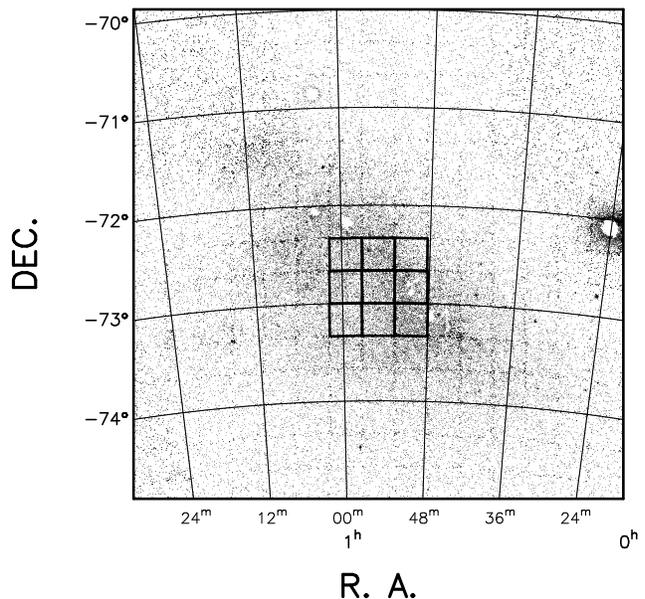}
\caption[]{A schematic map of the monitoring region in the SMC. Each mesh has a side of 20 arcmin and consists of 9 sub-meshes, which correspond to the field of view of the SIRIUS. The background stars are drawn from the GSC2.2 catalog. Stars brighter than 16th magnitude at F$_{\textrm{gsc}}$ located in the total area of 25 square degree centered on ($\alpha_{2000}$, $\delta_{2000}$) = (0$^{\textrm{h}}$ 55$^{\textrm{m}}$ 00.0$^{\textrm{s}}$, -70$^\circ$ 30$^\prime$ 00.0$^{\prime\prime}$) are plotted.}
\label{smc}
\end{figure}

We have been monitoring a total area of 3 square degrees along the LMC bar since December 2000, and a total area of 1 square degree around the SMC center since July 2001. Figure~\ref{smc} shows the monitoring area in the SMC. Details and some first results of the LMC survey are found in \citet{ita} (hereafter referred to as Paper I). In the present paper we deal with the single-epoch data from the monitoring survey.

\subsubsection{Photometry}
All data were reduced in the same way using the SIRIUS pipeline (Nakajima, private communication) and a single image comprises 100 dithered 5 sec exposures. Photometry was performed on the dithered images with DoPHOT in the fixed position mode (\citealt*{schechter}). The magnitudes are referred to the Las Campanas Ovservatory (LCO) system ones based on observations of a few dozen stars from \citet{persson}. In the system conversion, colour terms have been considered by observing red stars in table 3 of Persson et al. The photometry has a signal-to-noise ratio (S/N) of about 20 at 16.9, 16.8 and 16.0 mag in $J$, $H$ and $K$, respectively. Typical photometric errors are tabulated in table 1 of Paper I.

\subsubsection{Source selection}
\noindent \textbf{LMC:}\\
In the three square degree survey area, 1,593,248 sources were detected by DoPHOT. Because of the scan overlaps, there should be many multiple entries of a single source in the original sample. We eliminated such multiple entries based on (1) spatial proximity ($|\Delta r| \le 0.5^{\prime\prime}$) and (2) difference in photometry ($|\Delta m_{J,H and K}| \le 0.1$ mag). It leaves 1,305,983 sources. Further, requiring detections at least two of the three wave bands leaves 789,929 sources. 

In Paper I, we used the allowed discrepancy $|\Delta r|$ of $1.5^{\prime\prime}$, instead of the present value of $0.5^{\prime\prime}$. This was because the celestial coordinates were calculated by referencing the USNO catalog instead of the Guide Star Catalog II (GSC2.2, see section 2.1.3.). We can find substantial number of positional reference stars by using the GSC2.2 catalog, and this greatly improved the accuracy of the positioning.

\noindent \textbf{SMC:}\\
In the one square degree survey area, 257,637 sources were detected by DoPHOT. Multiple entries were eliminated in the same manner as we did in the LMC. It leaves 210,566 sources. Further, requiring detections at least two of the three wave bands leaves 107,376 sources.

\subsubsection{SIRIUS data}
In table~\ref{detection} we summarize the number of sources detected in each wave band. Hereafter, we refer these 789,929 and 107,376 sources as the SIRIUS data. Although many stars were detected only in two wave bands, these stars hardly affect the results of this paper. Also, although SIRIUS $JHK$ photometries are single-epoch ones, it should not affect the results so much, because the pulsation amplitudes of variable stars are usually small in the infrared (except for Mira variables, but still the results should be statistically fair because of large number of available data).

\begin{table}
\caption[]{Breakdown list of the SIRIUS data}
\label{detection}
\begin{center}
\begin{tabular}{cccccrr}
\hline
\multicolumn{5}{c}{} & \multicolumn{2}{c}{Number} \\
\multicolumn{5}{c}{Wavebands} & \multicolumn{1}{r}{LMC} & \multicolumn{1}{r}{SMC} \\
\hline
$J$&+&$H$&+&$K$& 596,913 & 85,491  \\
$J$&+&$H$& &   &  75,812 & 15,173  \\
   & &$H$&+&$K$& 114,827 &  6,693  \\
$J$& & + & &$K$&   2,377 &     19  \\
\hline
\multicolumn{5}{l}{Total} & \multicolumn{1}{r}{789,929} & \multicolumn{1}{r}{107,376} \\
\hline
\end{tabular}
\end{center}
\end{table}

The SIRIUS data were corrected for the interstellar absorption based on the relations in \citet{koornneef}, assuming $R =$ 3.2. We adopted ($A_J$, $A_H$, $A_K$) $=$ (0.129, 0.084, 0.040) and (0.082, 0.053,0.025), corresponding to the total mean reddenings of $E_{B-V} =$ 0.137 and 0.087 as derived by \citet{udalski1999a} for the OGLE's observing fields in the LMC and SMC, respectively.

\subsubsection{Astrometry}
Celestial coordinates of the sources detected by the SIRIUS were systematically calculated in the International Celestial Reference System (ICRS) by referencing the Guide Star Catalog II (\citealt{stsci}). The astrometric precision of the SIRIUS data is expected to be the same as that of GSC 2.2 catalog, i.e., better than 0.5$^{\prime\prime}$ in most cases (Tanab\'{e} et al. in preparation).

\subsection{OGLE}
Thanks to the microlensing surveys (OGLE, MACHO, EROS, MOA), a large number of variable stars were found as the natural by-products. In particular, the Optical Gravitational Lensing Experiment (OGLE) project has been monitoring the central parts of the Magellanic Clouds in the $BVI$ wave bands and the $I$ band time-series data (OGLE-II; \citealt{udalski}) is now available over the Internet (OGLE homepage; http://sirius.astrouw.edu.pl/$^{\sim}$ogle/).

\subsubsection{Cross-identification of the SIRIUS sources and the OGLE variables}
The SIRIUS data and all the stars found by the OGLE survey in the Magellanic Clouds have been cross-identified using a simple positional correlation. In the very strict sense, the OGLE survey does not fully cover the SIRIUS survey area, and some very small areas are not included in the OGLE. However, it never affect any conclusions in this paper.

\noindent \textbf{LMC:}\\
There are 49,008 OGLE variables in the OGLE/SIRIUS overlapped 3 square degree area in the LMC. The identification was made in two steps. First a search radius of $3^{\prime\prime}$ was used, finding 35,726 matches. Plotting the offsets between the SIRIUS coordinates and the OGLE coordinates revealed that there are small median astrometric shifts between them; the median differences are -0.80$^{\prime\prime}$ in right ascension (R.A.) and 0.10$^{\prime\prime}$ in declination (DEC). Second, the OGLE coordinates were corrected for these offsets, and a search radius of $3^{\prime\prime}$ was used again. We finally found 35,783 matches with an rms error in the difference between SIRIUS and OGLE coordinates 0.68$^{\prime\prime}$ in R.A. and 0.63$^{\prime\prime}$ in DEC, respectively. 

\noindent \textbf{SMC:}\\
There are 7,345 OGLE variables in the OGLE/SIRIUS overlapped 1 square degree area in the SMC. The identification was made in the same way as we did in the LMC. First a search radius of $3^{\prime\prime}$ was used, finding 6,118 matches. The offsets between the SIRIUS coordinates and the OGLE coordinates are 0.02$^{\prime\prime}$ in R.A. and -0.40$^{\prime\prime}$ in DEC. After correcting the OGLE coordinates for these median offsets, a search radius of $3^{\prime\prime}$ was used again. We finally found 6,103 matches with an rms error in the difference between SIRIUS and OGLE coordinates 0.41$^{\prime\prime}$ in R.A. and 0.38$^{\prime\prime}$ in DEC, respectively. 

\begin{figure}
\centering
\includegraphics[angle=-90,scale=0.355]{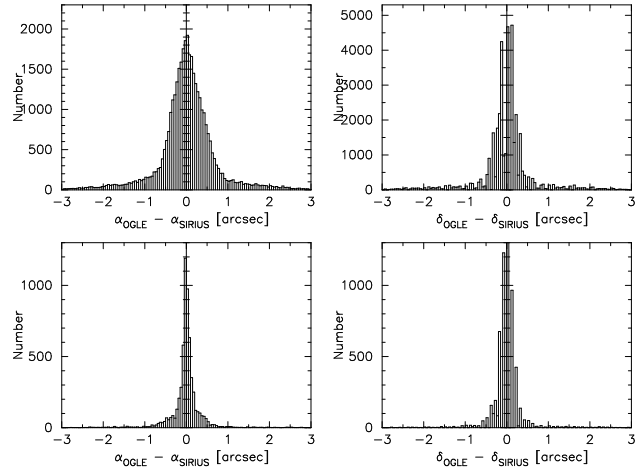}
\caption[]{Histogram of the positional difference between the SIRIUS coordinates and the OGLE coordinates in 0.05$^{\prime\prime}$ bins for the cross-identified stars in the LMC (upper panel) and the SMC (lower panel).}
\label{position}
\end{figure}

We show the histogram of positional differences of the finally identified stars in Figure~\ref{position}. It is clear that the positional differences are smaller than $1^{\prime\prime}$ for most of the identified stars (29,228 out of 35,783 stars or 81.7\% in the LMC, and 5,745 out of 6,103 stars or 94.1\% in the SMC).

\subsubsection{Period determination}
We applied the Phase Dispersion Minimization (PDM) technique to the OGLE data to determine the light variation periods of the cross-identified stars. This technique is described in detail by \citet{stelling} and here we show only the outline of the technique we used.

We represent the observational dataset as $(x_i, t_i)$, where $x$ is the magnitude, $t$ is the observation time and $i$ denotes the $i$th observation. We assume that there are $N$ observations in all ($i = 1, N$). Let $V$ be the variance of the magnitude of the original dataset, given by
\begin{equation}
V = \sum_{i=1}^N (x_i - \bar x)^2 / (N - 1),
\end{equation}
where $\bar{x}$ is the mean; $\sum x_i / N$. First, the $x_i$ and $t_i$ are folded with a trial period and the full phase space $[0, 1)$ is divided into an appropriate (user-specified) number of bins (there are $N_B$ bins in all, i.e., $j = 1, N_B$). Second, the variance $v_j$ of magnitude of each bin and the mean variance $\bar v$ of them are calculated, namely,
\begin{equation}
\bar v = \sum_{j=1}^{N_B} v_j / N_B.
\end{equation}
Third, the statistic parameter $\theta$, which is defined as
\begin{equation}
\theta = \bar v / V,
\end{equation}
is calculated for the trial period. These three processes are then repeated for the next trial period (i.e., $P_{\textrm{trial, new}} = P_{\textrm{trial, old}} + \textrm{resolution}$; see below). The $\theta$ is a measure of the regularity of the light variation or the scatter about the mean light curve, being hopefully near 0 for the very regular light variation with true period and near 1 for the irregular light variation without any distinctive periodicity or for the incorrect period. Hereafter, we call stars with small $\theta$ ($\leq$ 0.55) as ``regularly pulsating variables'', and the others ($>$ 0.55) as ``less regularly pulsating variables''.

We searched the minimum of the $\theta$ throughout the trial periods that start from 0.1 and end in 1,000 days. The trial periods were incremented by variable step values (= resolution), which are summarized in table~\ref{resolution}. For all of the 35,783 (LMC) and 6,103 (SMC) matched stars, their original light curve, $\theta$ spectra and the folded light curve are carefully eye-inspected, and the obvious spurious data points were eliminated. We assigned one pulsation period to each variable star, and the multi-periodic stars (e.g., \citealt{bedding}) were not analyzed in this paper. It should be noted that the total available baseline of the OGLE-II data is about 1,200 days and we did not try to determine the light variation period for the variables that seem to have the period longer than 1,000 days. Also, we did not try to determine the period for the variables that have too few good data points and that seem to hardly show any periodicity. The OGLE-II survey data has yearly blank phases that could be a problem for variable stars with pulsation period of about a year. However, experimentations showed that these yearly gaps hardly affect the period determination, and it is the virtue of the PDM technique. In very rare cases, the yearly gaps may unfortunately coincide with the timing of the peak luminosities. They can affect the determination of pulsation amplitude, but still, such cases should be very rare. Finally, we determined the light variation periods for 9,681 and 2,927 OGLE variables in the LMC and SMC, respectively.

\begin{table}
\caption[]{Used resolution in period search}
\label{resolution}
\begin{center}
\begin{tabular}{rrr}
\hline
\multicolumn{2}{c}{trial period [day]} & \\
\multicolumn{1}{r}{start($\ge$)} & \multicolumn{1}{r}{end($<$)} & \multicolumn{1}{r}{resolution [day]} \\
\hline
0.1 & 1.0 & 0.00001 \\
1.0 & 5.0 & 0.0001  \\
5.0 & 10.0 & 0.001 \\
10.0 & 100.0 & 0.01 \\
100.0 & 1000.0 & 0.1 \\
\hline
\end{tabular}
\end{center}
\end{table}

\section{Discussion}
\subsection{$K$ magnitude distribution}
Many low-amplitude variables were newly discovered by the aforementioned optical surveys. The vast majority are at the luminosities below the TRGB. \citet{alves} suggested that they are AGB stars in the early evolutionary phase prior to entering the thermal-pulsing phase of the AGB. In Paper I, we found that the $K$ magnitude distribution of the variable red giants in the LMC has two peaks; one is well above the TRGB, consisting genuine AGB variables and the other is just around the TRGB.

\begin{figure}
\centering
\includegraphics[angle=0,scale=0.455]{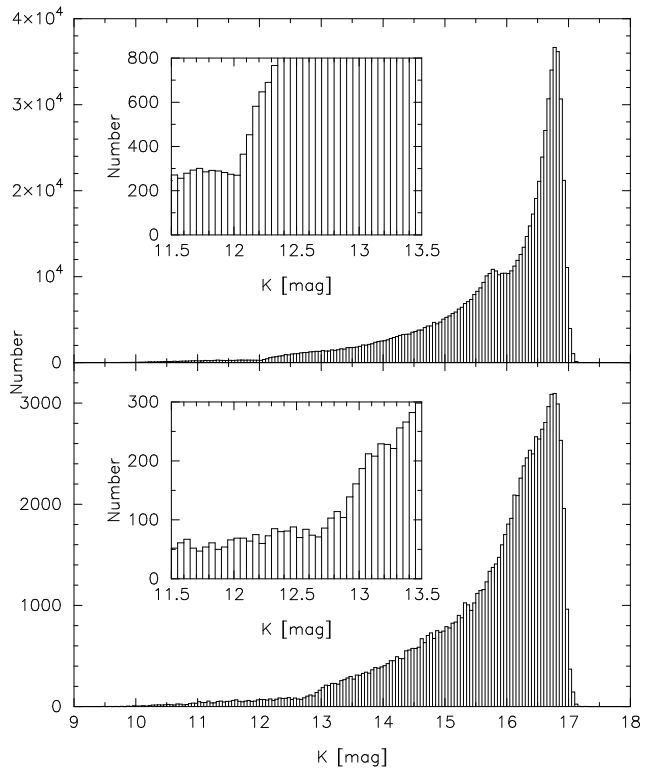}
\caption[]{The $K$ magnitude distribution differentiated by 0.05 mag bins of the SIRIUS data in the LMC (upper panel) and the SMC (lower panel). Closeup around the TRGB is shown in the inset.}
\label{lfall}
\end{figure}

\begin{figure}
\centering
\includegraphics[angle=0,scale=0.455]{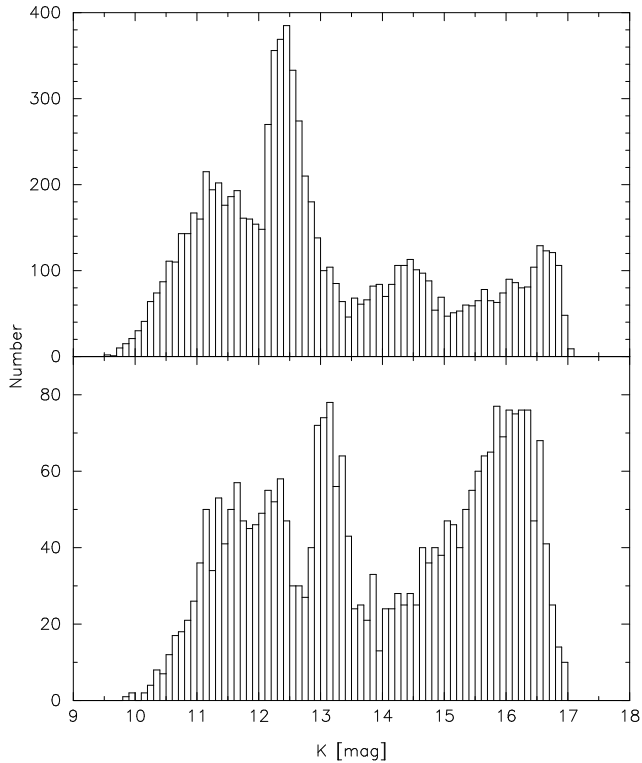}
\caption[]{The $K$ magnitude distribution differentiated by 0.1 mag bins of the OGLE/SIRIUS 9,681 and 2,927 variables in the LMC (upper panel) and the SMC (lower panel), respectively.}
\label{lfvar}
\end{figure}

In figure~\ref{lfall}, we show the $K$ magnitude distribution $N(K)$ ($dK =$ 0.05 mag) of the SIRIUS data. The discontinuity in the $N(K)$ is clearly seen around $K \approx$ 12.1 (upper panel) and $\approx$ 12.7 (lower panel), which corresponds to the TRGB of the LMC and SMC, respectively. Figure~\ref{lfvar} is the same diagram as the figure~\ref{lfall}, but the 9,681 and 2,927 OGLE/SIRIUS variables in the LMC and SMC are shown. It is clear that there is a ``pile-up'' around the $K$-band luminosity of the TRGB in both two galaxies. In Paper I, we concluded that a significant fraction of the constituent variable stars of the second peak could be on the RGB. Of course there may be faint AGB variables, but there is no reason to assume, according to models, that they accumulate at the TRGB.

\subsection{Period-$K$ magnitude diagram}
\citet{wood2000} found that radially pulsating red giants in the LMC form ``three'' distinctive parallel sequences in the $PK$ plane. He characterized the complex variability of the red giants by the pulsation modes, which are an important probe of the internal structure of the stars. \citet{wood1999} suggested that Mira variables are fundamental mode pulsators, while so-called semi-regular variables can be pulsating in the first, second or even higher overtone mode, by comparison of observations with theoretical models. However, a controversy still exists on the pulsation mode of Mira variables. Diameter measurements of Mira variables in the Milky Way indicates first overtone pulsation (e.g., \citealt*{haniff}, \citealt{whitelock2000}), while observations of radial velocity variation favor fundamental mode pulsation for Mira variables (e.g., \citealt*{bessell}).

\begin{figure}
\centering
\includegraphics[angle=0,scale=0.455]{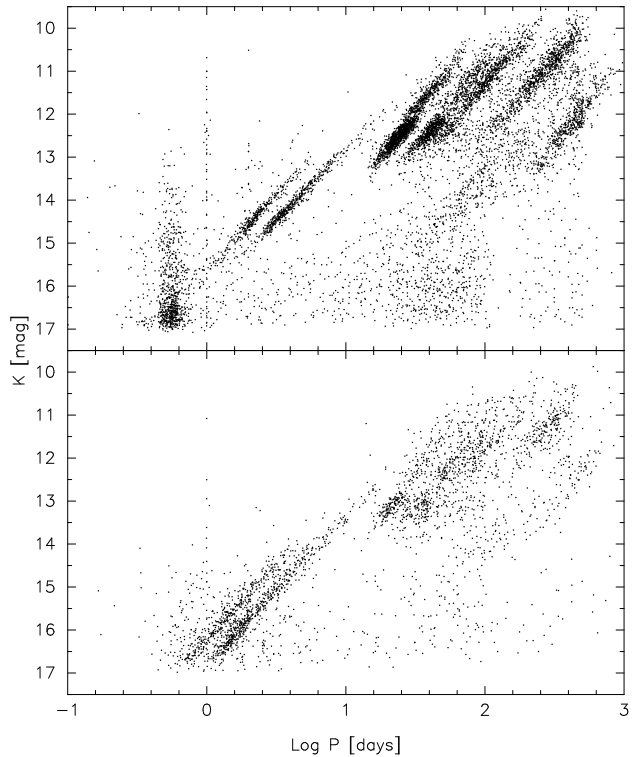}
\caption[]{Period-$K$ magnitude diagram for OGLE variables in the LMC (upper panel) and the SMC (lower panel).}
\label{plbw}
\end{figure}

We show the relationship between the determined pulsation periods and the $K$ magnitudes of the cross-identified variables in figure~\ref{plbw}. We found that the variable red giants in the SMC form parallel sequences on the $PK$ plane, just like those found by \citet{wood2000} in the LMC. In the LMC $PK$ diagram, bunch of variable stars exist in the faint magnitude ($K \lesssim$ 15.0) and very short period ($\log P \lesssim -0.1$) region. Considering their location on the $PK$ plane, they are likely to be RR Lyrae variables. Compared to the LMC $PK$ diagram, the SMC one is a bit scattered. This is probably due to the difference in the intrinsic depth of the two galaxies, because the LMC has a nearly face on orientation (\citealt{westerlund}), and \citet{groenewegen} obtained the total front-to-back depths of 0.043 mag for the LMC and 0.11 mag for the SMC, respectively. 

\begin{figure}
\centering
\includegraphics[angle=0,scale=0.455]{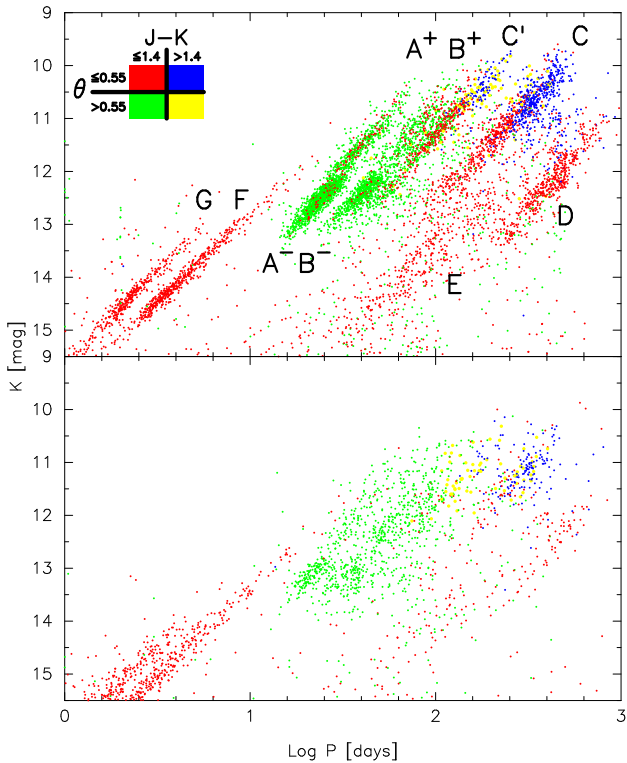}
\caption[]{Period-$K$ magnitude diagram for OGLE variables in the LMC (upper panel) and the SMC (lower panel). Variable stars are classified into four types according to their $J - K$ colours and $\theta$. The labels of the sequences are named in analogous to the sequences found by \citet{wood2000} in the LMC (except $C^\prime$, $F$ and $G$, see text).}
\label{plkct}
\end{figure}

In figure~\ref{plkct}, we again show the $PK$ relationship of variable stars in the Magellanic Clouds. In this diagram, we divided variable stars into four types according to their $J - K$ colours and $\theta$ as is shown on the upper-left corner of the figure. The labels of the sequences are named in analogy to the ones found in the LMC by \citet{wood2000} (except $C^\prime$, $F$ and $G$). While \citet{cioni2001} found no objects on sequence $A$ from their EROS/DENIS variables in the LMC, this sequence is clearly seen in both galaxies. This is essentially the same diagram as the one obtained by \citet{wood2000}. However, the sample used here is much larger than those of previous studies, and we can see new features in the figure: (1) the sequences $A$ and $B$ of \citet{wood2000} are found to be composed of the upper and lower sequences $A^+, B^+$, and $A^-, B^-$; (2) the sequence $B$ of \citet{wood2000} separates into three independent sequences $B^\pm$ and $C'$; (3) radially pulsating red giants form actually ``four'' parallel sequences ($A, B, C^\prime$ and $C$).

\subsubsection{Less regularly pulsating variables}
Most of the less regularly pulsating variables reside in the sequences $A$ and $B$, and show small pulsation amplitudes ($\Delta I \lesssim$ 0.2 mag, but some of the $B$ stars have $\Delta I$ of about 0.4 mag). Although it is not mentioned in \citet{wood2000}, his sequences $A$ and $B$ are densely populated at fainter magnitudes, and a discernible gap around the TRGB can be seen in his diagram (see his figure 1). This gap also presents in our diagram around the $K$-band luminosity of the TRGB ($K \sim$ 12.1 in the LMC and $\sim$ 12.7 in the SMC), which we roughly estimated in section 3.1. Also, in the LMC, a horizontal misalignment of the sequences apparently occurs at the TRGB. One possible explanation for the discontinuity is that neither sequences $A$ nor $B$ are single sequences, but each of them is composed of two independent sequences. So we subdivided the sequences $A$ and $B$ into $A^+, B^+$ and $A^-, B^-$, where the $+$ and $-$ correspond to the luminosity, being brighter than $K =$ 12.1 (LMC) or 12.7 (SMC) and fainter than it, respectively. 

Since the sequences $A^+$ and $B^+$ exceed the $K$-band luminosity of the TRGB, they consist of intermediate-age population and/or metal-rich old population AGB stars. However, the interpretation of the sequences $A^-$ and $B^-$ is difficult. If the metal-poor and old AGB stars pulsate in the same mode as the more massive stars on the sequences $A^+$ or $B^+$, they should be located on the sequences $A^-$ or $B^-$ because they do not exceed the TRGB luminosity. The question is whether the sequences $A^-$ and $B^-$ contain RGB pulsators or not. If AGB stars alone make up the $A^-$ and $B^-$ as well as $A^+$ and $B^+$ sequences, the small gap and the horizontal shift between them needs to be explained. If $A^-$ and $B^-$ stars, at least some of them, are pulsating on the RGB, then the discontinuity between the $-$ and $+$ sequences is understood in terms of different evolutionary stages. If we could apply the same period-temperature relation as that for Mira variables (\citealt{alvarez1997}) to the $A^-$ and $A^+$ stars, the period shift of $\delta \log P = -0.083$ between the two sequences at the TRGB ($K = 12.1$, see table~\ref{plrtable}) corresponds to the temperature difference of $\delta \log T_{\textrm{eff}} \approx 0.015$. This is in good agreement with the temperature difference between the AGB and RGB stars with the TRGB luminosity, predicted by the stellar evolutionary model for stars with 1$\sim$2 M$_\odot$ in the LMC (\citealt{castellani}). Though we prefer the RGB interpretation, it should be noted that a definitive answer has not been obtained.

\subsubsection{Regularly pulsating variables}
\begin{figure}
\centering
\includegraphics[angle=-90,scale=0.34]{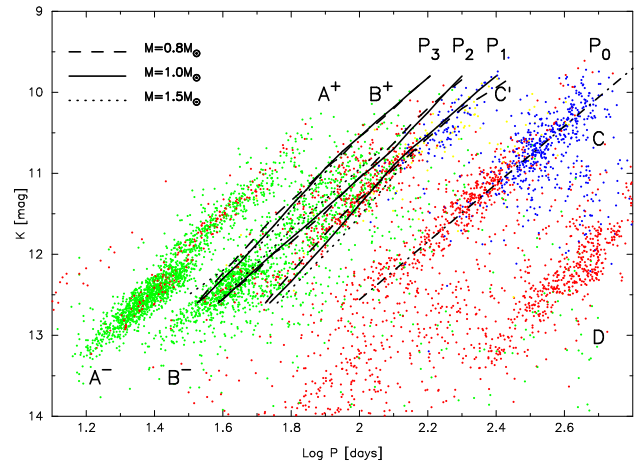}
\caption[]{Period-$K$ magnitude relations of variable red giants compared with theoretical model of \citet{wood1996}. Their models for stars of masses of 0.8, 1.0 and 1.5 M$_\odot$ pulsating in the fundamental (P$_0$, they forced theoretical fundamental mode period to fit the observed Mira relation of \citet{feast1989} (dot-dashed line)) and first (P$_1$), second (P$_2$) and third (P$_3$) overtone modes are shown. Colours of the points are as in figure~\ref{plkct}.}
\label{mode}
\end{figure}

Figure~\ref{mode} shows the comparison between the observational period-$K$ magnitude sequences of pulsating red giants in the LMC and the theoretical models calculated by \citet{wood1996}. The location of sequence $C$ is consistent with the LMC Mira sequence (\citealt{feast1989}). After classifying stars by their regularity of light variation, we discovered a new sequence $C^\prime$. Most of the stars on sequence $C^\prime$ are regularly pulsating just like those on $C$, but their amplitudes are smaller than those of the $C$ stars. We note, however, that the amplitudes of stars on sequence $C^\prime$ are larger than those of stars on $B$. The observed periods, luminosities and period ratio of sequences $C$ and $C^\prime$ agree very well with those of the theoretical fundamental (P$_0$) and first overtone mode (P$_1$) pulsation models (\citealt{wood1996}). Also, the theory predicts that overtone pulsators have smaller amplitudes than fundamental ones. Therefore, all of the above observational facts and the agreements with the theory could be naturally explained if we consider that the stars of sequences $C$ and $C^\prime$ are Mira variables pulsating in the fundamental and first overtone mode, respectively. However, the pulsation theory cannot reproduce the period ratio of the observed sequences $A$ and $B$. To explain their pulsation modes, a new theoretical model is needed.

\begin{figure}
\centering
\includegraphics[angle=0,scale=0.52]{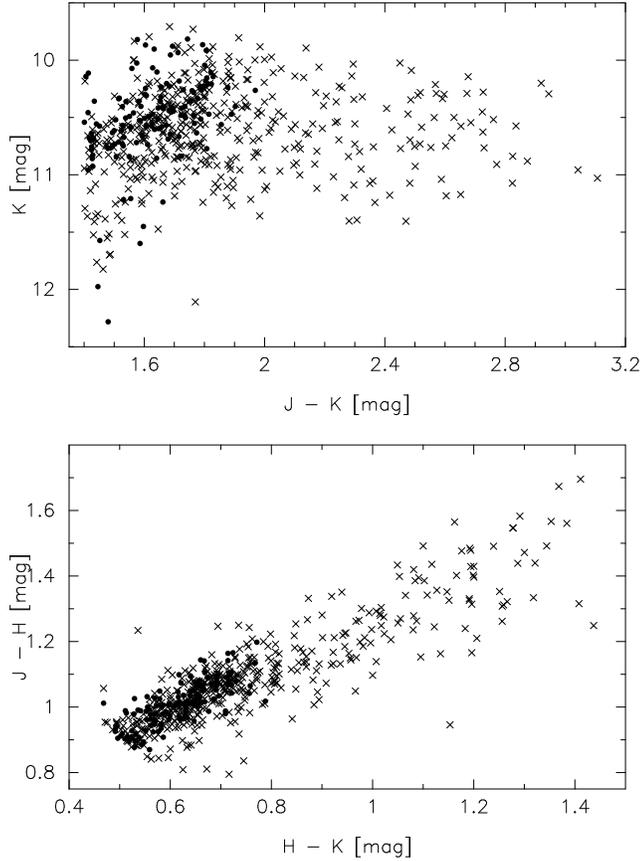}
\caption[]{Colour-magnitude (upper panel) and colour-colour (lower panel) diagram of the carbon-rich stars on sequence $C$ (crosses) and  $C^\prime$ (dots). Two stars that are on sequence $C$ are excluded from the colour-colour diagram, because they are outside the range.}
\label{carbon}
\end{figure}

It is interesting that, with only a few exceptions, all of the red giants with red colours ($J - K > 1.4$) reside in the brighter magnitudes of sequences $C$ and $C^\prime$. According to \citet{nikolaev}, stars with $J - K > 1.4$ in the LMC are primarily carbon-rich AGB stars, being considered as the descendants of oxygen-rich AGB stars. We subdivided stars on sequences $C$ and $C^\prime$ into two subgroups on the basis of their $J - K$ colour, being redder than 1.4 and the others. Table~\ref{plrtable} (see section 3.3) infers that those two populations may follow the different $PK$ relations, and the carbon-rich ($J - K > 1.4$) stars could define a line of somewhat shallower slope with brighter zero point than that of the oxygen-rich stars. \citet{feast1989} obtained $PK$ relations for carbon- and oxygen-rich stars in the LMC. The calculated slopes and zero-points are in good agreement with their results. In figure~\ref{carbon}, we plotted colour-magnitude and colour-colour diagram of probable carbon-rich stars on sequence $C$ and $C^\prime$. Interestingly, all of the carbon-rich stars with $J - K$ colour redder than 2.0 are on the sequence $C$, meaning that all of them are pulsating in the fundamental mode. However, it should be emphasized that the obscured oxygen-rich stars such as OH/IR stars could have redder colours. So, it could be dangerous to assume that all of the stars with redder colours are carbon-rich stars. Infrared spectroscopic work would provide the definite conclusion on these issues. 

The sequence $D$ is thought to be populated by the longer period of a binary system (\citealt{wood2000}). However, the explanation of this sequence is not clear yet. The periods of the group $D$ stars are very long and more long-term observations are needed for the clear explanation. \citet{wood1999} suggested that stars on the loose sequence $E$ are contact and semi-detached binaries. Interestingly, their distribution seems to extend to the TRGB. Stars on sequences $F$ and $G$ are very regularly pulsating and have periods ranging from less than 1 day to more than 30 days, suggesting that they are Cepheid variables pulsating in fundamental and first overtone mode, respectively.

\subsection{Period-$K$ magnitude relations}
\begin{figure}
\centering
\includegraphics[angle=0,scale=0.455]{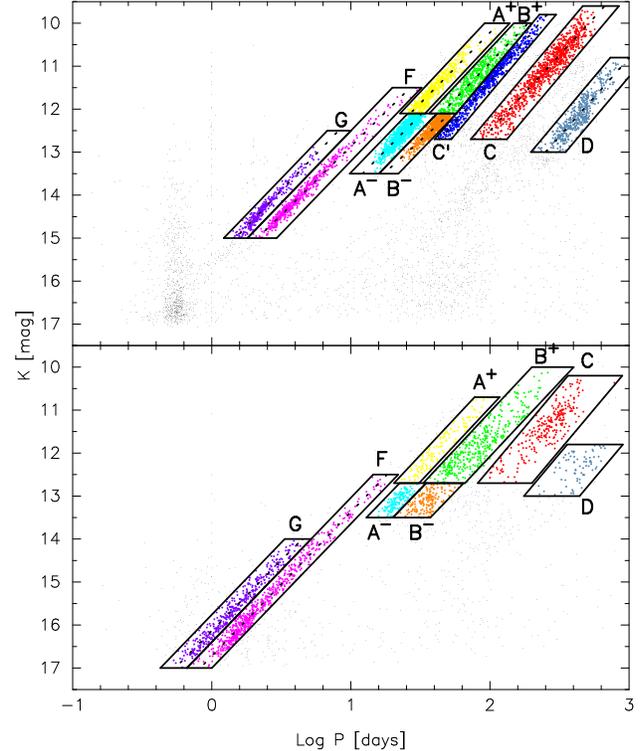}
\caption[]{Period-$K$ magnitude diagram for OGLE variables in the LMC (upper panel) and the SMC (lower panel) with classification boxes. The dotted lines are the least-square fits of linear relation to stars in each box.}
\label{pl}
\end{figure}

In order to determine the $PK$ relations, we defined boundary lines as in figure~\ref{pl} that divide the variable stars into nine (LMC) and eight (SMC) prominent groups (sequence $E$ is excluded). For groups $C$, $C^\prime$ and $D$ the slopes of the boundary lines were chosen to be -3.85, which is obtained by \citet{hughes} for the slope of the $PK$ relation for the blend populations of carbon- and oxygen-rich Miras in the LMC. The slope of the right side of group $B^+$ was also chosen to be -3.85. For groups $A^-$$B^-$, $A^+$$B^+$ and $F$$G$, the slopes of the boundary lines were chosen to be -3.35, -3.45 and -3.36, respectively, after the rough estimation of the slopes of each sequence. The width of the each box was chosen so that each of them contains the most probable body of a sequence and it minimizes the miss-classifications as much as possible.

\begin{table}
\caption[]{Period-$K$ magnitude relations for variable stars in the Magellanic Clouds of the form $K = a \times \log P + b$. The $K$ magnitudes are referred to the LCO system \citep{persson}. The $\sigma$ is the standard deviation in $K$, and $N$ is the number of stars being used in the least square fitting after applying 3.0 $\sigma$ clipping. Stars on sequences $C$ and $C^\prime$ in the LMC were subdivided into two subgroups according to their $J - K$ colour, and their period-$K$ magnitude relations were also calculated.}
\label{plrtable}
\centering
\begin{tabular}{lrrrrl}
\hline
\multicolumn{1}{l}{Group} & \multicolumn{1}{c}{$a$} & \multicolumn{1}{c}{$b$} & \multicolumn{1}{c}{$\sigma$} & \multicolumn{1}{c}{$N$} & \multicolumn{1}{l}{$J-K$}\\
\hline
\multicolumn{6}{c}{\textbf{LMC}}\\
\hline
$A^-$      & -3.284$\pm$0.047 & 17.060$\pm$0.065 & 0.114 &  1142 & All \\
$A^+$      & -3.289$\pm$0.047 & 16.793$\pm$0.077 & 0.125 &   510 & All \\
$B^-$      & -2.931$\pm$0.057 & 17.125$\pm$0.091 & 0.100 &   584 & All \\
$B^+$      & -3.356$\pm$0.052 & 17.634$\pm$0.099 & 0.160 &   502 & All \\
$C^\prime$ & -3.768$\pm$0.023 & 18.885$\pm$0.046 & 0.110 &   693 & All \\
$C^\prime$ & -3.682$\pm$0.109 & 18.720$\pm$0.245 & 0.122 &   135 & $>$1.4 \\
$C^\prime$ & -3.873$\pm$0.031 & 19.081$\pm$0.060 & 0.105 &   558 & $\le$1.4 \\
$C$        & -3.520$\pm$0.034 & 19.543$\pm$0.082 & 0.198 &   975 & All \\
$C$        & -3.369$\pm$0.099 & 19.165$\pm$0.250 & 0.200 &   447 & $>$1.4 \\
$C$        & -3.589$\pm$0.055 & 19.698$\pm$0.126 & 0.196 &   528 & $\le$1.4 \\
$D$        & -3.635$\pm$0.078 & 21.718$\pm$0.207 & 0.198 &   472 & All \\
$F$        & -3.188$\pm$0.019 & 16.051$\pm$0.013 & 0.095 &   540 & All \\
$G$        & -3.372$\pm$0.041 & 15.574$\pm$0.016 & 0.091 &   317 & All \\
\hline
\multicolumn{6}{c}{\textbf{SMC}}\\
\hline
$F$        & -3.350$\pm$0.020 & 16.722$\pm$0.010 & 0.139 &   606 & All \\
$G$        & -3.280$\pm$0.035 & 16.133$\pm$0.009 & 0.147 &   403 & All \\
\hline
\end{tabular}
\end{table}

Because the scattering is rather large due probably to the SMC's intrinsic front-to-back depth, we did not tried to separate $C^\prime$ stars from $B^+$ ones in the SMC. Also, we did not calculate the $PK$ relations of variable red giants in the SMC. The dotted lines in figure~\ref{pl} are the least square fits of linear relation to each group, and table~\ref{plrtable} summarizes the calculated $PK$ relations.

\subsubsection{Metallicity effects on the period-$K$ magnitude relations}
\citet{gascoigne} empirically obtained the period-visual magnitude relation for Cepheid variables and found that Cepheids in the metal deficient environment would be fainter than those with the same periods in the metal rich one. Based on this result, the Cepheids in the metal-deficient SMC should be fainter than those in the LMC if compared with the same period. On the other hand, \citet{wood1990} suggested that if there was a factor of 2 difference in metal abundance between LMC and SMC long period variables (e.g., Miras), the LMC stars were fainter by 0.13 magnitudes in $K$ than those in the SMC with the same period. Therefore, Cepheids and long period variables are expected to express completely different reactions to the metal abundance.

\begin{figure}
\centering
\includegraphics[angle=-90,scale=0.34]{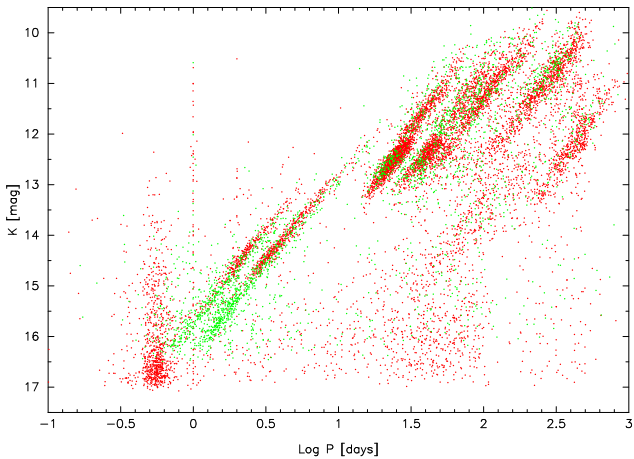}
\caption[]{Period-$K$ magnitude diagram of variable stars in the LMC (red dots) and SMC (green dots). The $K$ magnitudes of SMC stars are shifted by $-0.49$ magnitude (see text) to account for the difference in the distance moduli between the LMC and the SMC.}
\label{metal}
\end{figure}

One possible way to test this anticipation is to compare the $PK$ relations in the LMC and SMC, and it is done in figure~\ref{metal}. The distances to the Magellanic Clouds are still uncertain, and we can only say with confidence that the SMC lies some 0.4-0.6 mag beyond the LMC \citep{cole}. We adopted the distance moduli to the LMC and SMC as 18.40 (\citealt{nelson2000}) and 18.89 (\citealt*{harries}), respectively. These distance moduli were obtained by using the eclipsing binary systems in the Magellanic Clouds, and the technique, while not entirely geometrical (reddening estimates are required), is free from the metallicity-induced zero point uncertainties. So, we shifted the $K$ magnitudes of the SMC stars by $-0.49$ magnitude to account for the different distance moduli of the SMC and LMC. The red dots represent the LMC stars and the green dots correspond to the SMC stars.

The figure shows, just as expected, that Cepheids in the SMC are fainter, but red giants in the SMC are brighter than those in the LMC if compared with the same period. Also, it is likely that the slopes of the $PK$ relations are almost the same in the two galaxies and only the zero points seem to differ. To estimate the differences quantitatively, we fixed the slopes of the $PK$ relations of SMC Cepheids and Miras to the corresponding LMC ones, and calculated the zero points. The fixed-slope solutions yielded the zero points of 19.409$\pm$0.018, 16.170$\pm$0.006 and 15.659$\pm$0.007 mag for sequences $C$, $F$ and $G$ in the SMC, respectively (after shifting their $K$ magnitudes by $-0.49$ mag). Comparing these zero points with those of the LMC Cepheids and Miras, Cepheids in the SMC are fainter by $\approx 0.1$ mag and Miras in the SMC are brighter by $\approx 0.13$ mag than the LMC ones with the same period. 

\section{Summary}
We cross-identified the OGLE-II survey data and the SIRIUS NIR data in the Magellanic Clouds. Based on the large and homogeneous data\footnote{Data of the variable stars will be published in the separate paper.}, we studied the pulsation properties of variable stars in the Magellanic Clouds. We discovered a new sequence on the period-$K$ magnitude plane, and identified it as the one for Mira variables pulsating in the first overtone mode. We also found that the variable red giants in the SMC form parallel sequences in the period-$K$ magnitude plane, just like those found by \citet{wood2000} in the LMC. The period-$K$ magnitude relations were derived, especially for both fundamental mode and overtone Cepheids in the Magellanic Clouds. These results are obtained with the greatest numbers of available stars and most deep photometric data in the infrared up to now. Metallicity effects on period-$K$ magnitude relations are studied by comparing variable stars in the Magellanic Clouds. The result suggested that Cepheids in the metal-deficient SMC are fainter than the LMC Cepheids, while Miras in the SMC are brighter than the LMC ones if compared with the same period.

\section*{Acknowledgments}
The authors thank the referee, Dr. Albert Zijlstra for constructive comments. It is a pleasure to thank Dr. Michael Feast for valuable and helpful comments on the first version of this paper. We also thank Dr. Ian Glass for useful comments. This research is supported in part by the Grant-in-Aid for Scientific Research (C) No. 12640234 and Grant-in-Aids for Scientific on Priority Area (A) No. 12021202 and 13011202 from the Ministry of Education, Science, Sports and Culture of Japan. The IRSF/SIRIUS project was initiated and supported by Nagoya University, National Astronomical Observatory of Japan and University of Tokyo in collaboration with South African Astronomical Observatory under a financial support of Grant-in-Aid for Scientific Research on Priority Area (A) No. 10147207 of the Ministry of Education, Culture, Sports, Science, and Technology of Japan. The Guide Star Catalogue-II is a joint project of the Space Telescope Science Institute and the Osservatorio Astronomico di Torino. Space Telescope Science Institute is operated by the Association of Universities for Research in Astronomy, for the National Aeronautics and Space Administration under contract NAS5-26555. The participation of the Osservatorio Astronomico di Torino is supported by the Italian Council for Research in Astronomy. Additional support is provided by  European Southern Observatory, Space Telescope European Coordinating Facility, the International GEMINI project and the European Space Agency Astrophysics Division.


\label{lastpage}

\end{document}